\documentclass[prb,twocolumn,letter,floatfix,longbibliography]{revtex4-1}
\usepackage{graphicx}

\usepackage{amsmath}
\usepackage{comment}
\usepackage[dvipsnames]{xcolor}

\begin{document}

\title{Strong Particle-Hole Symmetry Breaking\\
in a 200 kelvin Superconductor} 
\author{Soham S. Ghosh$^{1}$}
\author{Yundi Quan$^{1}$}
\author{ Warren E. Pickett*$^{1}$}
\affiliation{
$^1$ Department of Physics, University of California Davis, Davis, California 95616, USA\\
}
\date{\today}
\begin{abstract}
{\bf 
The superconducting state of metals has long provided a classic example of particle-hole symmetry (PHS) at low energy. Fermionic self-energy results based on first principles theory for the electron-phonon coupling in H$_3$S presented here illustrate strong PHS-breaking dynamics arising from the underlying sharp structure in the fermionic density of states. Thus H$_3$S is not only the superconductor with the highest critical temperature $T_c$ (through 2018), but its low energy, low temperature properties deviate strongly from textbook behavior.  The minor momentum and band dependence of the fermionic self-energy allows evaluation of the momentum-resolved and zone-averaged spectral densities and interacting thermal distribution function, all of which clearly illustrate strong particle-hole asymmetry.
 }
 \end{abstract}
 \maketitle
 
 \vskip 2cm



{\bf Background.} The discovery\cite{drozdov2015} and confirmation\cite{h3sexp2,h3sDuan2017} of record high superconducting transition temperature as high as T$_c$=203 K in H$_3$S at 160 GPa pressure has re-ignited efforts toward the long dreamt of pinnacle of superconductivity at room temperature. The conventional phonon-exchange mechanism of electron pairing confirmed by isotope shift measurements\cite{drozdov2015} and by several theoretical studies\cite{duan2014,bernstein2015,papa2015,flores2016,heil2015,akashi2015,errea2015,ge2016} belies the occurrence and interconnection of several {\it unconventional} processes competing within H$_3$S.  This new phase of superconducting matter above 200K arises from three causes: the small mass of the proton compared to that of typical nuclei, strong scattering by H displacements, and sharp structure in the electronic density of states $N(E)$ at the Fermi energy ($E_F$) arising from a pair of closely spaced (300 meV) van Hove singularities (vHs). Each vHs occurs at 24 symmetry related points in the Brillouin zone (BZ), thereby involving a significant fraction of the Fermi surface with low carrier velocities.\cite{yundi2016} The former aspect elicits phenomena that have received detailed theoretical attention: high frequency phonons; quantum zero point motion\cite{akashi2015} of H and its effect on crystal stability;\cite{errea2016} strong anharmonicity, which impacts dynamic crystal stability;\cite{errea2016,akashi2015} the character and coupling strength of phonons;\cite{duan2014,heil2015,akashi2015,errea2016} and the unsettled question of non-adiabatic effects due to vanishing velocities near E$_F$ at 48 points in the BZ.\cite{yundi2016}


\begin{figure*}[htp]
\vskip 0.2 in
\begin{center}
\includegraphics[width=0.95\textwidth]{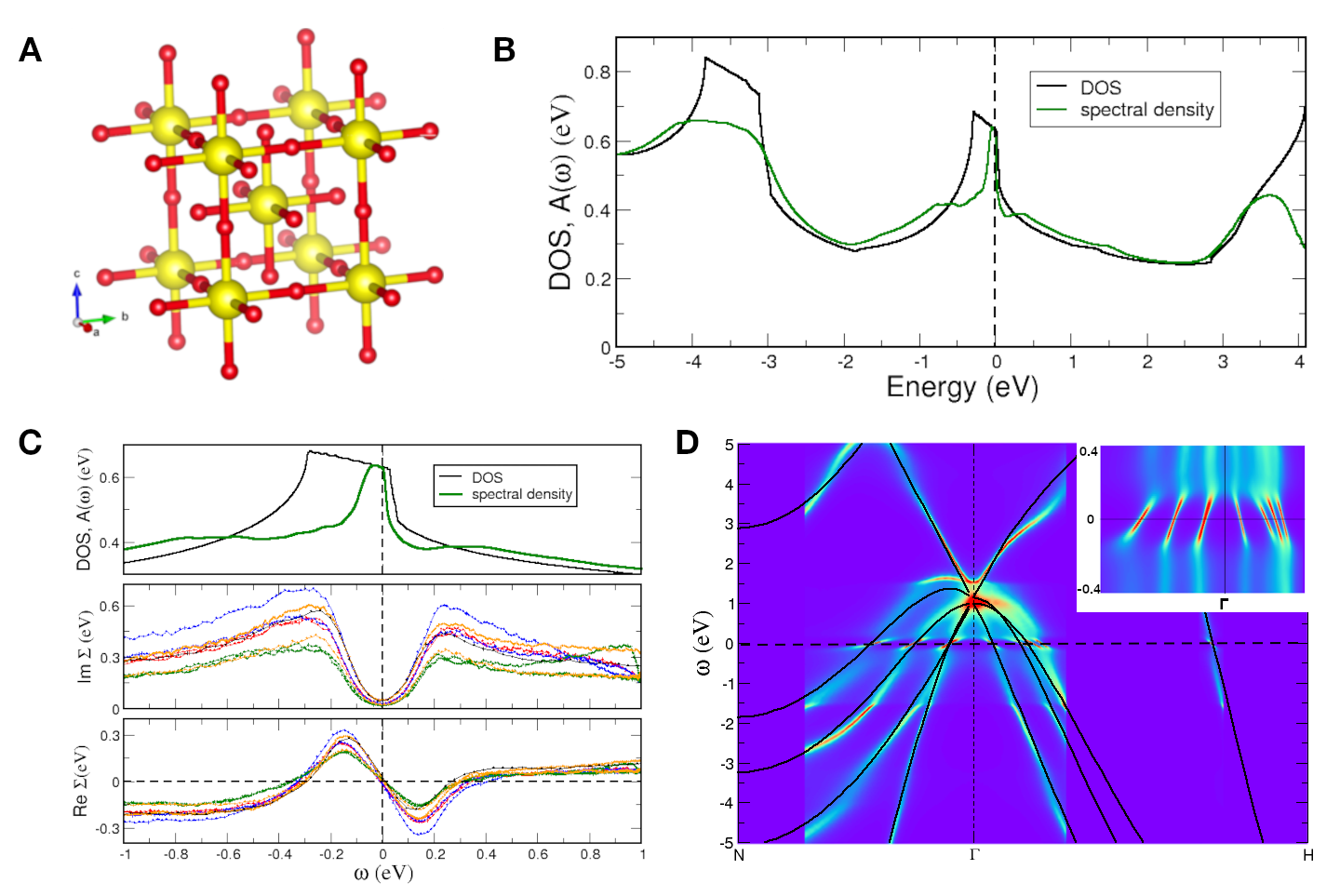}
\end{center}
\caption{ 
         \textbf{Electron self-energy $\mathbf{\Sigma_{k,n}(\omega)}$ and several consequences.}
         (\textbf{A}) Crystal structure of $Im\bar3 m$ H$_{3}$S with highlighted H-S bonds, large and small spheres 
         represent S and H, respectively. Two S atoms form a simple body centered cubic lattice and 
         H atoms are located between two S atoms.
         (\textbf{B}) Spectral density $A(\omega)$ (see text), compared to the electronic DOS. Note the sharpness of the peak at E$_{F}$.
         (\textbf{C}) Bottom panel: $Re \Sigma_{k,n}(\omega)$, for the five bands ($n$) crossing $E_F$ plotted versus the frequency $\omega$. Middle panel: $Im \Sigma_{k,n}(\omega)$, obtained and plotted as for $Re \Sigma_{k,n}(\omega)$. Top panel: the behaviour of spectral density $A(\omega)$ and the electronic DOS near the E$_{F}$ peak, showing the distinctive {\it narrowing} rather than broadening of the peak.
         (\textbf{D}) Band ($n$) and momentum ($\vec k$) resolved electron spectral function $A_{\vec k,n}(\omega)$ calculated with EPW, overlaid on the uncoupled Kohn-Sham bands, along two high symmetry lines. Colors at the red end of the rainbow spectrum are higher intensity. The renormalized slope at the Fermi level (see inset) is an indication of the factor of three (1+$\lambda$) mass enhancement caused by strong EPC. Further than $\Omega \sim$ 250 meV from the Fermi level, decay by phonon emission smears the single particle states and removes the coherent renormalization.
         }
\label{fig:fig1}
\vskip 0.2 in
\end{figure*}


    The effects of the vHs are neglected in most theoretical work, although every aspect of standard electron phonon coupling (EPC) theory requires reformulation when $N(E)$ varies strongly on the scale of phonon energies.\cite{wep1982} Momentum-dependent Eliashberg equations can be applied on the imaginary frequency axis\cite{wep1982,akashi2015} but the vHs make numerical convergence challenging. Additionally, the normally irrelevant Debye-Waller self-energy (SE) bubble diagram becomes relevant.\cite{akashi2015} Also, the concept of phonon scattering electrons from Fermi surface to Fermi surface must be generalized because regions of the zone with energy within $\Omega$ (maximum phonon energy) of the Fermi surface contribute non-uniformly to scattering processes. In the case when only the density of states(DOS) but not the wavefunction character is strongly varying, as in H$_3$S,\cite{papa2015,yundi2016} the theory has been generalized and was applied to Nb$_3$Sn,\cite{wep1982} whose DOS peak is even sharper than in H$_3$S.   

There has not yet been any study of the interacting fermionic excitation spectrum in H$_3$S that provides direct evidence of EPC. The pioneering work of Engelsberg and Schrieffer\cite{engelsberg1963} and Shimojima and Ichimura\cite{shimojima1970} revealed how coupling results in mixed elementary excitations -- fermionic quasiparticles and renormalized (bosonic) phonons -- that are each combinations of bare electrons and phonons, with their mixing increasing with the EPC strength $\lambda$. In this paper we address the excitation spectrum, with description of (1) {\it sharpening,not broadening} of the already narrow spectral peak by interactions, (2) the ``waterfall'' structure in the momentum-resolved spectral density as coherence burns off, and (3) several unconventional impacts of strong p-h symmetry breaking.

\vskip 3mm
{\bf Methods.}
Density functional theory based and density functional perturbation theory based computations
have been carried out using a plane wave basis set in the {\sc quantum espresso}\cite{QE-2009,QE-2017} (QE) simulation package, obtaining the electronic and phonon dispersions and EPC. H$_3$S 
at a pressure of $210$ GPa has space group $Im\bar3 m$ consisting of a bcc lattice with a lattice constant of $5.6$ a.u.\cite{papa2015}, pictured in Fig. 1(a). This structure and pressure has been used in the present calculations in harmonic approximation. Norm-conserving pseudopotentials of the Trouiller-Martins type are used in the QE code. The Perdew-Burke-Ernzerhof (PBE)\cite{perdew1996} implementation of the generalized gradient approximation (GGA) is chosen as the exchange-correlation functional. 

Bardeen-Cooper-Schrieffer (BCS) theory, and its strong coupling Migdal-Eliashberg extension,\cite{scalapino1966,giustino2017} of phonon-coupled superconductivity provides a broad roadmap for achieving high critical temperature in conventional superconductors.
Quantities related to EPC, such as the Eliashberg function $\alpha^2F(\omega)$, $\lambda$, the electron self-energy $\Sigma(\vec k,\omega)$, and resulting spectral functions in the normal state were obtained using the QE, Wannier90,\cite{mostofi2008} and EPW codes.\cite{giustino2007,ponce2016,margine2013,giustino2017} We have found that the vHs require unusually fine $\vec k$ (electron) meshes to obtain converged results. Along with a planewave basis set with a cutoff of 85 Ry we have used a $26\times26\times26$ \textbf{k}-point grid for the electronic self-consistent routines and a $32\times32 \times32$ \textbf{k}-point grid for the DOS calculation. A $6\times6\times6$ \textbf{q}-point grid was used to calculate the phonon eigenvalues and dynamical matrices. The use of Wannier90 and EPW allow us to interpolate the electron and phonon eigenvalues to a fine mesh of $60\times60\times60$ $\mathbf{k}$- and $\mathbf{q}$-points each for the final Brillouin zone (BZ) integration. The band electronic spectrum $N(E)$ contains a narrow peak with Fermi energy (the zero of energy) lying at the higher of two vHs separated by 300 meV,\cite{yundi2016} shown on different scales in Figs. 1(b) and (c).

The EPW\cite{giustino2007,ponce2016} code  of Giustino and collaborators provides the electron momentum $k$ and band index $n$ self-energies on a dense BZ mesh.  The weak $k,n$ dependence of the electronic self-energy $\Sigma_{kn}(\omega)$ allows construction of a zone-averaged $\Sigma(\omega)$ that will provide the interacting electronic spectral function. The fermionic Green's function $G_{k,n}(\omega)$ provides the band electron spectral density $A_{k,n}(\omega)$ and thereby the zone-averaged counterpart $A(\omega)$.

\vskip 3mm
{\bf Electron Self-energy.}

Scatter plots of both real and imaginary parts of $\Sigma_{kn}(\varepsilon_{kn})$ versus $\varepsilon_{kn}$ are shown in Fig. 1(c) for the bands crossing $\mu$ (zero). The self-energies fall along lines for each band, with minor band dependence. This enables us to obtain a representative average $\Sigma(\omega)$. 

The electron self-energy $\Sigma_{kn}(\omega)=M+i\Gamma$ gives the energy shift $M$ and broadening $\Gamma$ that provide insight into many consequences of EPC.  The interacting Green's function is (suppressing all indices) $G^{-1} = G_{KS}^{-1} -\Sigma$ in terms of the Kohn-Sham (band) Green's function $G_{KS}$.  In Fig.~\ref{fig:fig1}(c) these real and imaginary parts of $\Sigma_{kn}(\varepsilon_{kn})$, obtained using the EPW code in the normal state (200K) of H$_{3}$S, are shown for each of the five bands crossing E$_F$.  The EPW code produces matrix elements of $\Sigma$ in the DFT band representation $\Sigma_{k,n}(\omega)$, for each band state $\vec k, n$ at $\omega=\varepsilon_{kn}$. 

The band dependence is roughly an average value $\pm 30$\% in both $M$ and $\Gamma$, up to the characteristic frequency $\Omega$ of 250 meV (also roughly the maximum phonon frequency) where both $M(\omega)$ and $\Gamma(\omega)$ have begun to deviate from their low T, low energy Fermi liquid forms $M(\omega)=-\lambda \omega, \Gamma(\omega)=B(T)+C\lambda\omega^2$ for some constant $C$, and we use $B(200K)$=7 meV. The slope $-\partial M(\omega)/\partial \omega|_{\omega=0} = \lambda$ gives the quasiparticle mass enhancement $m^*/m_b=1+\lambda \sim 3.5$ over the band mass $m_b$. The inverse scattering rate $\Gamma$ peaks at a value of 0.6 eV at $|\omega|\approx 1.5\Omega$. Above 2$\Omega$ both parts of $\Sigma$ become slowly varying, with $M(\omega<-2\Omega) \approx$-0.2 eV for occupied states, $\Gamma(|\omega|>2\Omega) \approx$0.3 eV. Note that above $|\omega|\approx$ 2$\Omega$ $M$ is not particle-hole symmetric.

%
\vskip 3mm
{\bf Interacting band structure.}
The band-resolved spectral density $A_{kn}(\omega)\equiv|Im G_{kn}(\omega)|/\pi$, which is the interacting counterpart of the Kohn-Sham band structure and the quantity measured in angle-resolved photoemission probes, is shown as a heat map in Fig.~\ref{fig:fig1}(d).  This plot, with an enlargement of the low energy region in the inset, reveals several impacts of EPC. (1) The mass enhancement (flattening) of the sharp, low energy quasiparticle bands by the factor $1+\lambda$ at the chemical potential $\mu$=0. (2) The coherent mass enhancement ``boils away'' abruptly just above $\omega \sim \Omega$, giving way at higher energies to a substantial 0.3-0.5 eV broadening around spectral density peaks that are displaced by $\sim$0.2 eV from the Kohn-Sham bands. (3) The transition region reveals the ``knee'' or ``waterfall'' transition from coherent to incoherent that is a signature of strong EPC. This knee has received much attention in cuprate HTS.\cite{waterfall} 

\vskip 3mm
{\bf Spectral density.}
The zone-averaged electron spectral density is given by $A(\omega)= \frac{1}{\pi}\sum_{k,n}|Im G_{k,n}(\omega)|.$ We approximate by using a $k,n$-averaged self-energy $\Sigma(\omega)$, essentially equal to the red band data plotted in Fig.~\ref{fig:fig1}(c). 
[For small $\omega$ the self-energy presented by Kudryashov {\it et al.}\cite{kudryashov2017} is similar, since both satisfy the Fermi liquid conditions mentioned above. For larger $\omega$ there are strong differences due to approximations made in their work (see Supplemental Material).] 

The spectral function calculation simplifies to 
\begin{eqnarray}
A(\omega;\mu)=\int \frac{dE~~~N(E)~|\Gamma(\omega)|/\pi}
   {[\omega-(E-\mu)-M(\omega)]^2+[\Gamma(\omega)]^2}.
   \label{eq:spectral}
\end{eqnarray}
 This expression has the form of a Lorentzian broadening of $N(E)$, but with an energy shift $M(\omega)$ and broadening $\Gamma(\omega)$ that varies with $\omega$.  For comparison with experiment, the chemical potential $\mu=\mu(T)$ would need to be shifted to enforce conservation of particle number. Near the chemical potential where $\Gamma$ becomes very small, this expression reduces to a $\delta$-function and the behavior is $A(\omega)\approx N([1+\lambda]\omega)$. Thus the energy variation is enhanced by $1+\lambda \approx 3.5$, which results in the {\it strong narrowing} of the peak in $A(\omega)$ compared to that in in $N(E)$; this often unrecognized aspect is evident in Fig.~1(b).
 
 When $N(E)$ is constant over an energy range of a several $\Omega$ as in most metals and as assumed in textbooks, redistribution of spectral density is transparent to most probes. With sharp, asymmetric structure in $N(E)$ on the scale of $\Omega$ the effect is dramatic. Due to the Fermi liquid requirements listed above,
 Eq.~\ref{eq:spectral} gives $A(\omega=0;\mu)=N(E=\mu)$ which is reproduced in the calculation as shown in Fig. 1(c) and Fig. 2.

\begin{figure}[htp]
\begin{center}
\includegraphics[width=0.48\textwidth]{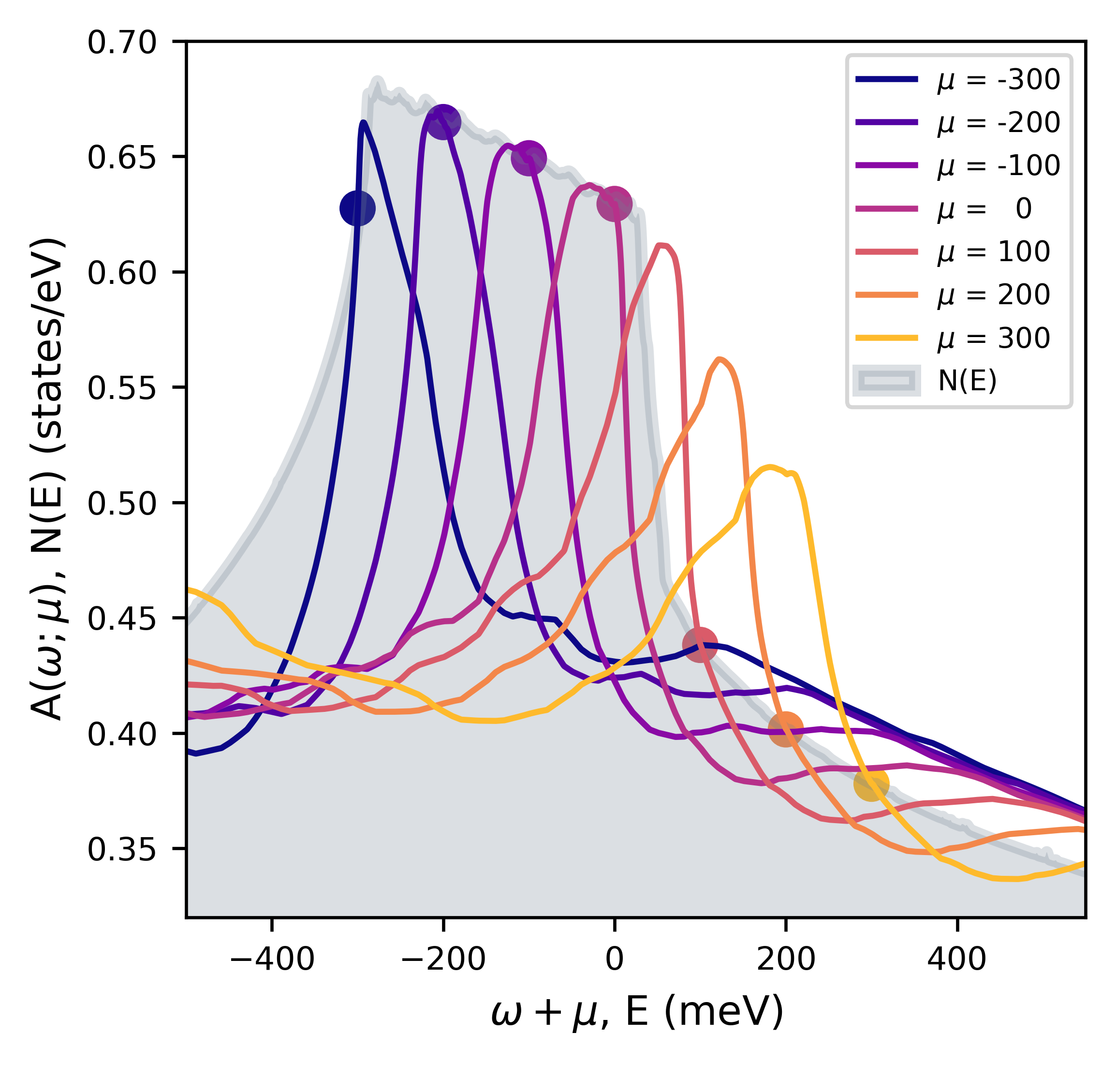}
\end{center}
\caption{
{\bf Doping dependence of spectral density.}
Dependence of the spectral density $A(\omega)$ on chemical potential $\mu$, calculated from Eq.~\ref{eq:spectral} at 200K. The circular dot on each graph marks the analytic low temperature constraint $A(\omega=0;\mu)=N(\mu)$, and the shaded region denotes N(E).  The strong particle- ($\mu > 0$) hole- ($\mu<0$) asymmetry is reflected in the simple translation of the peak for hole doping, while for particle doping the peak structure become highly asymmetric around $\mu$.
}
\label{spectral-mu}
\vskip 0.2 in
\end{figure}

We compare in Fig.~\ref{spectral-mu}, referenced to the underlying electronic DOS, the spectral density calculated for several band fillings at chemical potential $\mu$, neglecting changes in $\Sigma(\omega)$. 
The very strong asymmetry around the physical band filling of H$_3$S is evident. For zero or negative values of $\mu$, the spectral density remains a strongly compressed version of $N(E)$ and is displaced downward rigidly with $\mu$ across the plateau of $N(E)$. For electron doping, however, as $\mu$ passes across the upper vHs onto the steep slope, $A(\omega;\mu)$ peaks not at $\omega=0$ but at negative $\omega$. Doping ($\mu$) in effects drags the spectral density peak along with it for up to and beyond $\Omega$ above the intrinsic chemical potential. 
This asymmetry will induce behaviors of thermodynamic, transport, and low energy spectra that are quite different from what would be obtained from the bare spectrum, with the differences arising from the strong p-h {\it asymmetry}.

\begin{figure}[htp]
\vskip 0.2 in
\begin{center}
\includegraphics[width=0.48\textwidth]{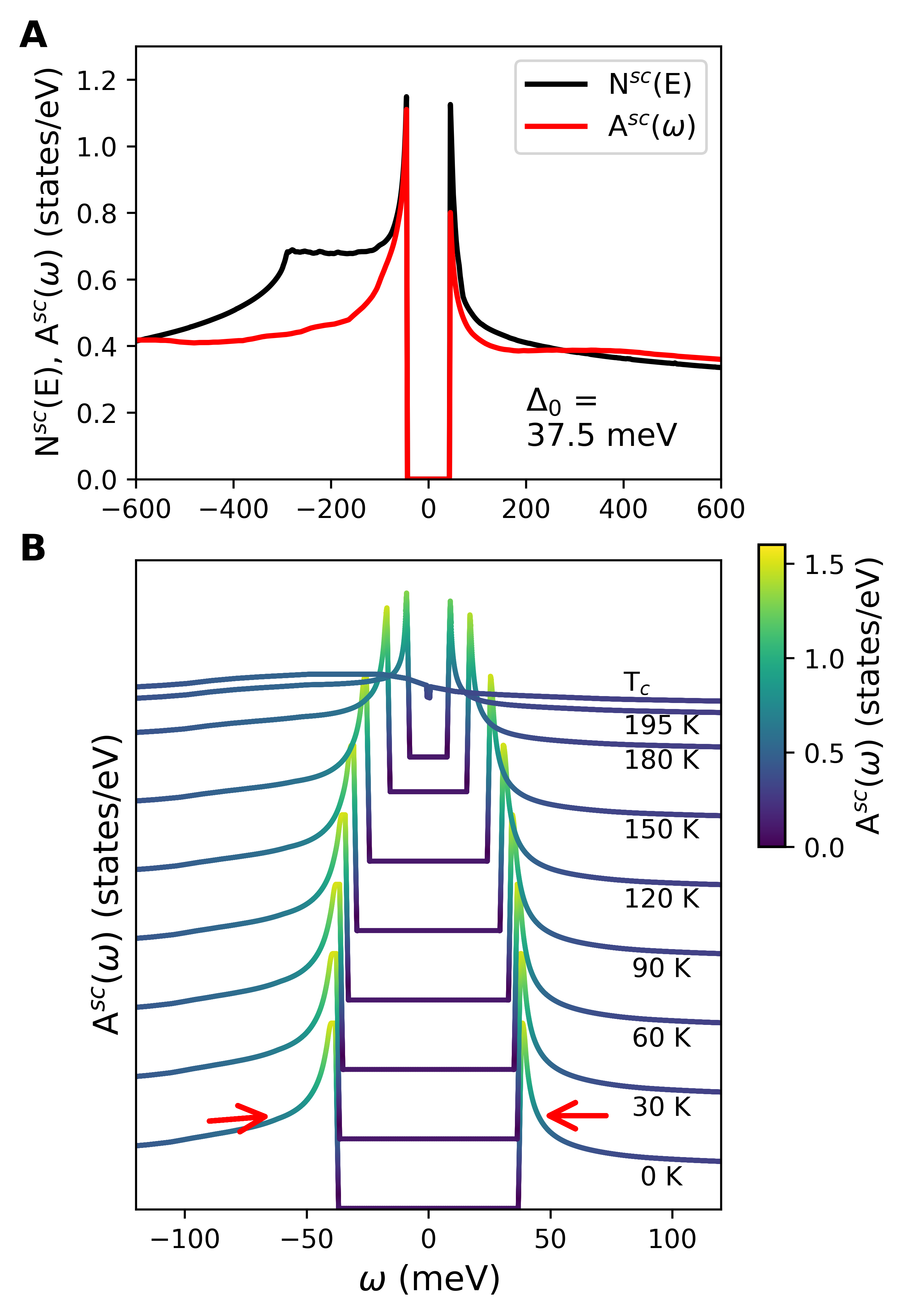}
\end{center}
\caption{{\bf Spectral density of H$_3$S in the superconducting state.} (A) Comparison at T=0 without (black) and with (red) the self-energy included.  
The strong particle-hole asymmetry across the gap is evident. The gap $2\Delta$=75 meV is the measured value\cite{capitani2017}.
(B) Temperature evolution of the spectral density $A^{sc}(\omega)$ through the T=0-200 K range, with constant vertical shifts of the curves. The color bar denotes the spectral intensity and the red arrows point to $0.8$ eV$^{-1}$ on either side of the gap, highlighting the particle-hole asymmetry.
}
\label{gap}
\end{figure}

\vskip 3mm
{\bf Effect of the Superconducting Gap.}
Careful treatment of T-dependences would require real-frequency-axis solutions of 
generalized Eliashberg equations (treating N(E)\cite{wep1982}), which are not available. Instead, standard BCS expressions can be adopted to allow demonstration of the implications of the vHs on the electronic spectral density. In Fig.~\ref{gap} we show the impact on $N(E)$ of opening of a superconducting BCS gap
\begin{eqnarray}
A^{sc}(\omega) = \int d\omega' A(\omega')\frac{\omega'}{\sqrt{(\omega'-\omega)^2-\Delta^2}}
\end{eqnarray}
with an analogous gap-opening expression for $N^{sc}(E)$ (with $\omega'\rightarrow E)$. The temperature dependence of the gap follows the approximate analytic form $2\Delta(T)=2\Delta_o[1-(T/T_c)^2]^{1/2}$, with $2\Delta_o \approx 5k_B T_c=75$ meV.  
Fig.~\ref{gap}(a) illustrates the great difference between $N_{sc}(E)$ and $A_{sc}(\omega)$, each displaying the strong particle-hole asymmetry imparted by the vHs. 

\vskip 0.2 in

Figure~\ref{gap}(b) displays the dependence on temperature of the quasiparticle spectrum $A_{sc}(\omega)$. The temperature dependence of $\Sigma(\omega)$ can be neglected since T$_c$, although at a record high value, is still low compared to phonon frequencies The gap remains sharp and the spectrum remains asymmetric until it assumes its normal state form at T$_c$.

\vskip 3mm
{\bf Effects on thermal distribution.}
When $\Sigma_{kn}(\omega)$ depends on $k,n$ only through $E_{kn}$ as is reasonably assumed for H$_3$S, (with the same holding for $A_{kn}(\omega)$) it is simple to show\cite{pickettprl1982} that for certain thermodynamic properties, the effect of interactions can be transferred to the thermal distribution function $f(\omega)$, the interacting counterpart of the non-interacting Fermi-Dirac distribution $f_o(E)$.  In Matsubara space, the definition in terms of the Green's function leads to ($k_B = 1 = \hbar$ units, $\mu$=0)
\begin{eqnarray}
f(E_{kn},T;\mu)&\equiv& T\sum_{n=-\infty}^{+\infty} G(kn,i\omega_n;\mu)e^{i\omega_n\eta} \nonumber \nonumber \\
&=&\int d\omega f_{0}(\omega)|\frac{1}{\pi}Im G(kn,\omega;\mu)|\label{eq:FD}
         \\
&\rightarrow&\int d\omega \frac{ f_{0}(\omega)|\Gamma(\omega)|/\pi}
   {[\omega-E_{kn}-M(\omega)]^2+[\Gamma(\omega)]^2}\nonumber
\label{eq:FD},
\end{eqnarray}
upon continuing back to the real frequency axis and then using the $k$-averaged, hence $k$-independent, self-energy.
In this equation $\omega_n = (2n+1)\pi T$ is the Matsubara frequency, and $\eta$ is a positive infinitesimal.

\begin{figure}[htp]
\begin{center}
\includegraphics[width=0.48\textwidth]{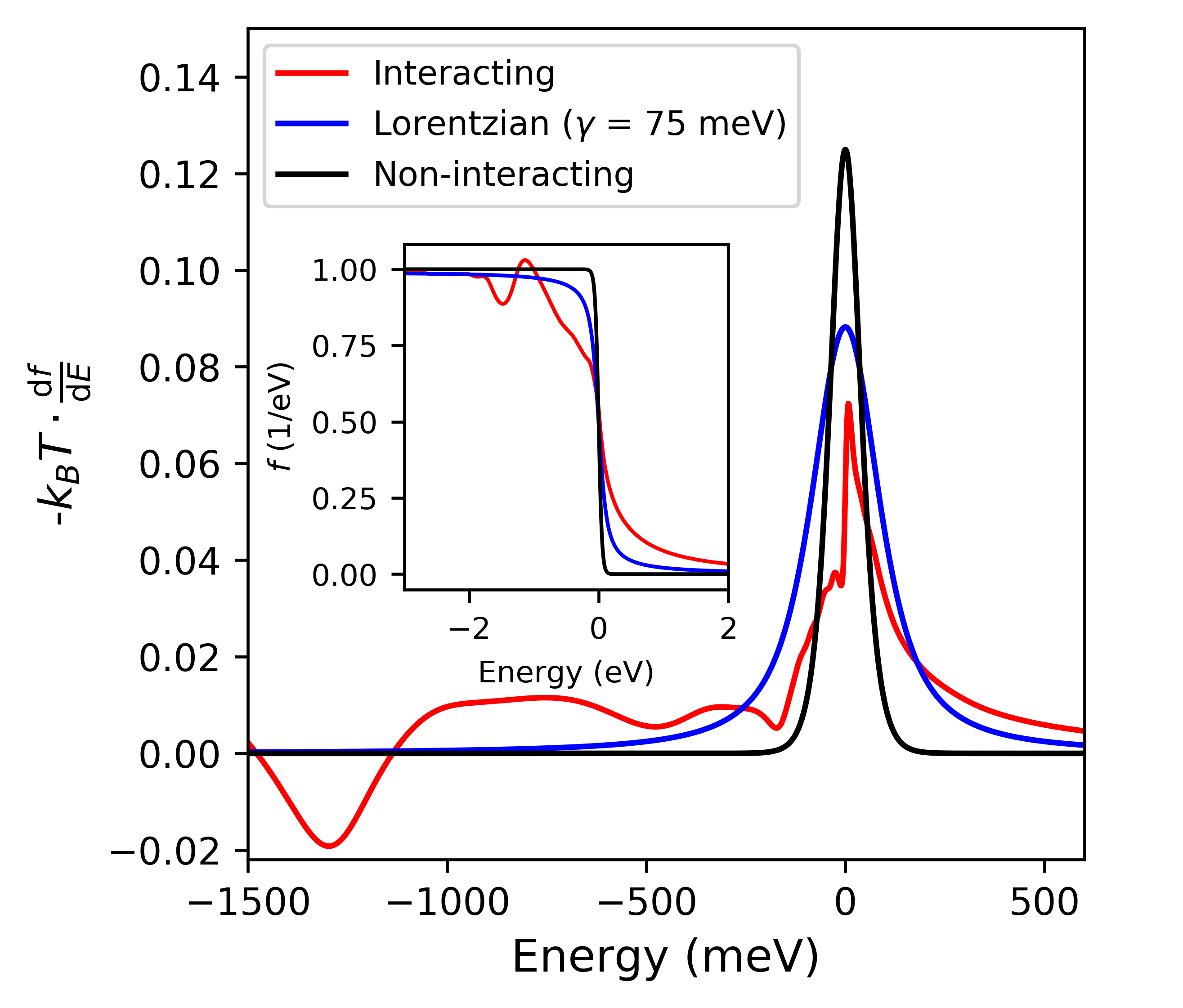}
\end{center}
\caption{
{\bf Interacting thermal distribution.}
 The dimensionless derivative of the thermal occupation -T$\frac{\partial f(E-\mu,T;\mu)}{\partial T}$  (red line), 
 contrasted with the non-interacting Fermi-Dirac, divided by two to fit on the same scale (black line). Also shown, to emphasize the difference in shape, 
 (blue line) is an interacting thermal distribution calculated using a fictitious spectral density of a Lorentzian form with half-width at half maximum of 75 meV. 
 Interaction results not only in the expected broadened of $f(E;\mu)$, but also strong asymmetry in particle-hole thermal occupation emerges. 
        The inset shows the broadening the distribution function itself. The distributions are calculated at $T=300$ K.
}
\label{thermal-distrib}
\vskip 0.2 in
\end{figure}

Along with the calculated spectral density, this allows one to write the electron number in three different ways, beginning with the non-interacting Kohn-Sham electron number ${\cal N}$ from the first expression ($E_F$=0):
\begin{eqnarray}
{\cal N}&=& \int dE f_{0}(E;T=0) N(E) \nonumber \\
 &=&\int d\omega f_{0}(\omega-\mu;T) A(\omega;\mu)\label{eq:electron_1} \\
 &=& \int dE f(E-\mu,T;\mu) N(E).
 \label{eq:electron_2}
\end{eqnarray}
The second relation determines $\mu(T)$, an estimation of which is provided in the supplementary material.
The second and third expressions together express that, for (conserved) particle number, the effect of interactions on ${\cal N}$ can be shifted between an interacting density of states $A(\omega;\mu)$ or an interacting thermal distribution function $f(E-\mu,T;\mu)$.
Eq.~\ref{eq:electron_2} is a simple example of a general result obtained by Lee and Yang\cite{leeyang1960}, that thermodynamic quantities can be obtained from thermal averages incorporating the interacting (versus non-interacting) occupation number, but in momentum space ({\it i.e.} $f(\vec k)$ versus $f_o(\varepsilon_{\vec k})$).

This interacting thermal distribution function reflects how non-interacting states [$N(E)$] are sampled by the interacting system, rather than the conventional formulation in which interacting kernels are sampled over the non-interacting thermal distribution.  Figure~\ref{thermal-distrib} provides the derivative 
$-T\frac{\partial f(E,T;\mu)} {\partial E}$. The interacting distribution differs strongly in the region of strong variation of $N(E)$. Unlike the symmetric derivative of the Fermi-Dirac distribution, the interacting version becomes asymmetric according to the behavior of $\Sigma(E)$ convoluted with the structure imposed by the vHs. Contrary to intuition, with interactions the thermal occupation is not limited to the interval [0,1], though excursions outside this range will have limited effect. 

The primary effect, shown in Fig.~\ref{thermal-distrib}, is significant redistribution 
leading to a strong distortion from particle-hole symmetry. To illustrate the point that the spectral function has a form qualitatively different from a Lorentzian broadening, we also show in Fig.~\ref{thermal-distrib} a convolution of the non-interacting
Fermi-Dirac distribution with a Lorentzian of halfwidth $\gamma$=75 meV.
This model (and p-h symmetric) distribution lacks the qualitative features of the real interacting distribution.

\vskip 3mm
{\bf Infrared spectrum.} Infrared spectroscopy is an important probe of the effects of EPC in superconductors. While a general study of the effect of vHs awaits further developments, something can be said about the far infrared where interband transitions are negligible. Capitani {\it et al.}\cite{capitani2017} have reported the observed and modeled reflectivity in H$_3$S at high pressure, observing structure identified with an energy scale of 160 meV, suggested to be a phonon with a very strong infrared coupling that is not evident from current theory. Given the paucity of experiments that are possible at high pressure, this experiment assumes unusually strong impact. 

\begin{figure}[!ht]
\includegraphics[width=0.48\textwidth]{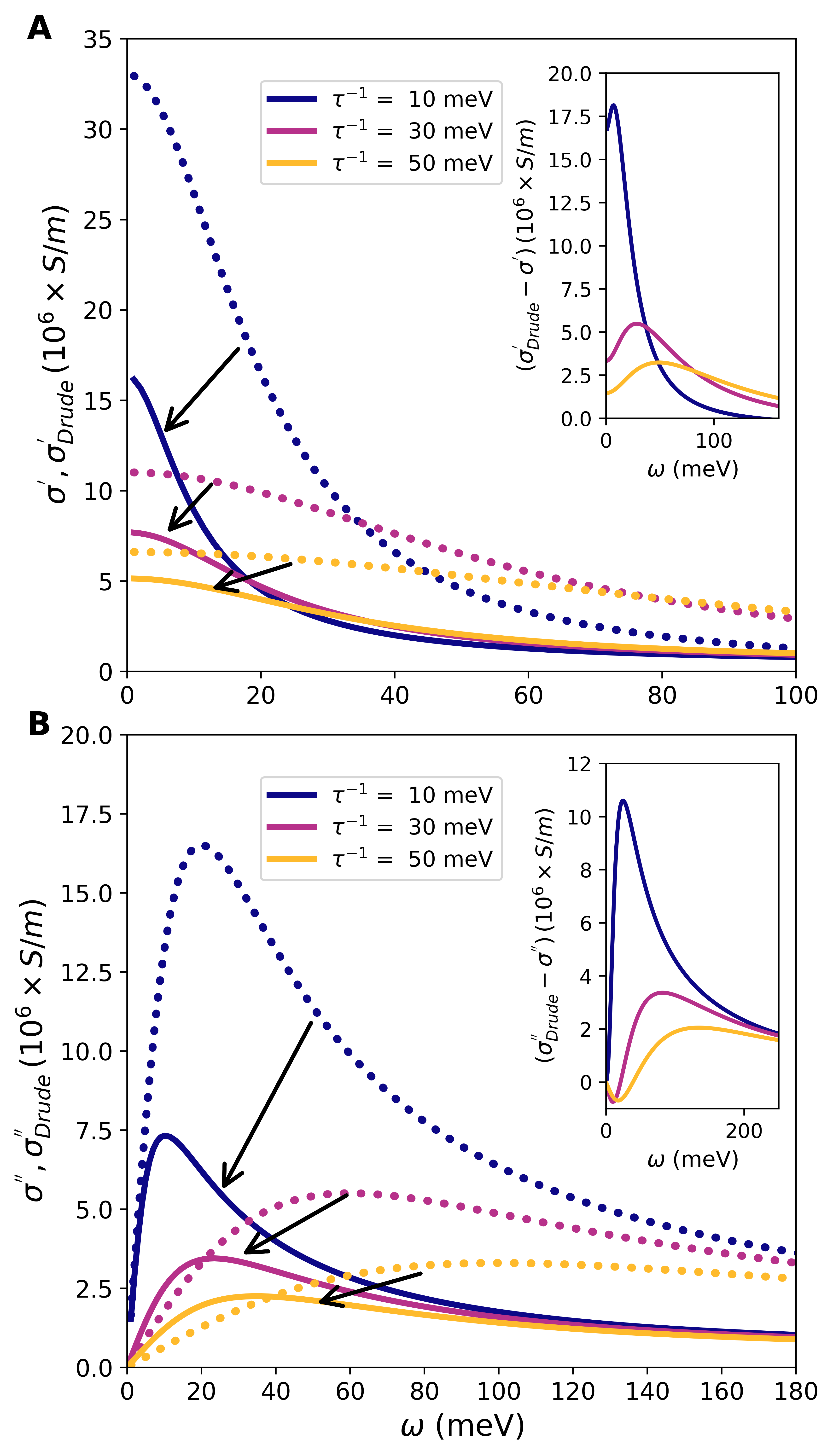}
\caption{{\bf Infrared optical conductivity of H$_{3}$S.}
(A) Real, and (B)  imaginary parts of conductivity in the normal-state H$_{3}$S, comparing the generalized Drude expression of Eq.~\ref{eq:allendrude} (solid lines) with the non-interacting [$\Sigma(\omega)=0]$ limit(dotted lines), at $T=200$ K. Results are shown for three values of the elastic scattering rate $1/\tau$. The insets show the differences as indicated.
}
\label{optical}
\end{figure}

Without EPC, the intraband optical conductivity reduces to the Drude form $\sigma(\omega)=\sigma_o/(1-i\omega\tau)$, where 
\begin{eqnarray}
\sigma_o = e^2 N(0)\frac{v_F^2}{3}\tau = e^2 \left(\frac{n}{m}\right)_{eff}\tau\equiv \frac{\omega_p^2\tau}{4\pi}
\label{eq:drude}
\end{eqnarray}
gives the zero frequency limit in terms of N(0) and the mean Fermi surface velocity\cite{bernstein2015} $v_F=2.5\times 10^7$ cm/s, which combines with $N(0)$ to give the Drude plasma frequency $\omega_p$=2.8 eV for H$_3$S. In the general expression for the current-current correlation function that gives $\sigma(\omega)$, the vertex matrix elements are simply the electron velocity $v_k$. In the average over the Fermi surface, the velocity factors cancel the contributions from the vHs regions, making the ubiquitous ``constant DOS'' treatment reasonable. Allen has given the corresponding generalized Drude (intraband) expression\cite{AllenDrude} for an interacting system 
\begin{eqnarray}
\sigma(\omega)=i\frac{\omega_p^2}{4\pi}
  \int d\omega'\frac{[f(\omega')-f(\omega' +\omega)]/\omega}
      {\omega -\Sigma(\omega' +\omega)+
              \Sigma^*(\omega')}.
\label{eq:allendrude}
\end{eqnarray}
The integral gives an effective $\tau_{eff}(\omega)$. Results with additional elastic scattering values of $1/\tau$=10, 30, 50 meV, are displayed in Fig.~\ref{optical}.
Inelastic scattering due to EPC causes a reduction (shown by the arrows) in the real component of the conductivity as it approaches its static dc value at small $\omega$. The $(1+\lambda)$ mass enhancement shows up in the imaginary part, where it translates to an increased slope at the low-frequency linear regime as well as a strong reduction in the magnitude. 
These results do not resolve the origin of the feature studied by Capitani {\it et al.}\cite{capitani2017}

\vskip 3mm

{\bf Summary.}
We have demonstrated the significant impact of strong electron-phonon coupling on the electron spectral density and band renormalization of the high-temperature superconductor H$_{3}$S, pointing out its relationship with the sharp and asymmetric DOS peak near the Fermi level caused by two neighboring vHs. The picture of a simple smeared spectral density is conceptually incorrect; the peak becomes sharper than the DOS at the Fermi level due to the mass enhancement effect. Peculiarly, the sharp spectral density peak follows the chemical potential with doping, but in a strongly particle-hole asymmetric way. In the superconducting regime, strong particle-hole symmetry breaking occurs as a consequence the van Hove singularities. We further elucidate the effect of electron-phonon coupling on the infrared conductivity, for consideration in interpretation of optical data. 

\newpage


\end{document}